\documentclass[useAMS,usenatbib,a4paper]{aa} 
\usepackage[varg]{txfonts}
\usepackage{url}
\usepackage{graphicx}
\usepackage{gensymb} 
\usepackage[nolist]{acronym}
\usepackage{siunitx}
\begin{acronym}[NANOGrav]
	\acro{ism}[ISM]{interstellar medium}
	\acro{iism}[IISM]{ionised interstellar medium}
	\acro{dm}[DM]{dispersion measure}
	\acro{toa}[ToA]{time of arrival}
	\acroplural{toa}[ToAs]{times of arrival}
	\acro{msp}[MSP]{millisecond pulsar}
	\acro{pta}[PTA]{pulsar timing array}
	\acro{epta}[EPTA]{European Pulsar Timing Array}
	\acro{ppta}[PPTA]{Parkes Pulsar Timing Array}
	\acro{nanograv}[NANOGrav]{North American Nanohertz Observatory for Gravitational Waves}
	\acro{ipta}[IPTA]{International Pulsar Timing Array}
	\acro{lofar}[LOFAR]{LOw Frequency ARray}
	\acro{ilt}[ILT]{International \ac{lofar} Telescope}
	\acro{hba}[HBA]{high-band antenna}
	\acro{lba}[LBA]{low-band antenna}
	\acro{glow}[GLOW]{German LOng Wavelength}
	\acro{mpifr}[MPIfR]{Max-Planck Institut f\"ur Radioastronomie}
	\acro{lump}[LuMP]{\ac{lofar} und \ac{mpifr} Pulsare}
	\acro{rfi}[RFI]{radio frequency interference}
	\acro{snr}[S/N]{signal-to-noise ratio}
	\acro{pulp}[PulP]{the pulsar pipeline}
\end{acronym}

\usepackage{dcolumn}
\newcolumntype{.}{D{.}{.}{-1}}
\newcommand{\dcol}[1]{\multicolumn{1}{.}{#1}}

\newcommand{\maxspan}{7.1}
\newcommand{\maxglowspan}{6.4}
\newcommand{\maxcorespan}{7.1}
\newcommand{\npsr}{36}

\newcommand{\npsrsolart}{12}
\newcommand{\npsrglowt}{six}
\newcommand{\npsrcoret}{18}

\newcommand{\firstglowobsdate}{20 August 2013}
\newcommand{\lastglowobsdate}{08 January 2020}
\newcommand{\firstcoreobsdate}{19 December 2012}
\newcommand{\lastcoreobsdate}{14 January 2020}
\newcommand{\nchanmin}{6}
\newcommand{\nchanmax}{20}

\newcommand{\meanprej}{0.7}
\newcommand{\solarpsrlisttext}{J0030+0451, J0034$-$0534, J0407+1607, J0645+5158, J1012+5307, J1022+1001, J1300+1240, J1400$-$1438, J1730$-$2304, J1744$-$1134, J2145$-$0750, and J2317+1439}
\newcommand{\solarpsrlowestelat}{J1022+1001}
\newcommand{\solarpsrlowestelatv}{0.1}
\newcommand{\solarpsrhighestelat}{J1012+5307}
\newcommand{\solarpsrhighestelatv}{38.8}
\newcommand{\solarnondetectionlisttext}{J0751+1807, J2051$-$0827, and J2222$-$0137}
\newcommand{\meanrfifrac}{9.0}
\newcommand{\ravgwindowlowlist}{J0030+0451, J0034$-$0534, J1022+1001, J1400$-$1431, J2145$-$0750, and J2317+1439}

\bibpunct{(}{)}{;}{a}{}{,} 

\newcommand{\DMunits}{cm$^{-3}$\,pc}
\newcommand{\tempoII}{\textsc{tempo}\oldstylenums{2}}

\usepackage{color} 

\begin{document}

\title{Dispersion measure variability for \npsr{}
  millisecond pulsars at 150\,MHz with LOFAR}

\author{
	J.~Y.~Donner\inst{\ref{mpi},\ref{ubi}}
	\and
        J.~P.~W.~Verbiest\inst{\ref{ubi},\ref{mpi}} 
        \and
        C.~Tiburzi\inst{\ref{astron}} 
        \and
        S.~Os\l owski\inst{\ref{swin1},\ref{swin2}} 
        \and
        J.~Künsemöller\inst{\ref{ubi}} 
        \and
        A.-S.~Bak~Nielsen\inst{\ref{mpi},\ref{ubi}} 
        \and
        J.-M.~Grießmeier\inst{\ref{orleans},\ref{nancay}} 
        \and
        M.~Serylak\inst{\ref{skasa},\ref{uniwesterncape}} 
        \and
        M.~Kramer\inst{\ref{mpi},\ref{manchester}} 
        \and
        J.~M.~Anderson\inst{\ref{tuberlin},\ref{gfz}} 
        \and
        O.~Wucknitz\inst{\ref{mpi}} 
        \and
        E.~Keane\inst{\ref{skajb}} 
        \and
        V.~Kondratiev\inst{\ref{astron},\ref{lebedev}} 
        \and
        C.~Sobey\inst{\ref{csiro}} 
        \and
        J.~W.~McKee\inst{\ref{toronto}} 
        \and
        A.~V.~Bilous\inst{\ref{uniamsterdam}} 
        \and
        R.~P.~Breton\inst{\ref{manchester}} 
        \and
        M.~Brüggen\inst{\ref{hamburg}}
        \and
        B.~Ciardi\inst{\ref{garching}}
        \and
        M.~Hoeft\inst{\ref{tautenburg}}
        \and
        J.~van~Leeuwen\inst{\ref{astron},\ref{uniamsterdam}} 
        \and
        C.~Vocks\inst{\ref{potsdam}}
}
\institute{
	Max-Planck-Institut f\"ur Radioastronomie, Auf dem H\"ugel 69,
	53121 Bonn, Germany
        \label{mpi}
	\and
	Fakult\"at f\"ur Physik, Universit\"at Bielefeld, Postfach 100131,
	33501 Bielefeld, Germany
        \label{ubi}
        \and
        ASTRON, the Netherlands Institute for Radio Astronomy,
        Oude Hoogeveensedijk 4, 7991 PD Dwingeloo, The Netherlands
        \label{astron}
        \and
        Gravitational Wave Data Centre, Swinburne University of Technology,
        P.O. Box 218, Hawthorn, VIC 3122, Australia
        \label{swin1}
        \and
        \label{swin2}
        Centre for Astrophysics and Supercomputing, Swinburne University of Technology,
        P.O. Box 218, Hawthorn, VIC 3122, Australia
        \and
        \label{orleans}
        LPC2E – Université d’Orléans / CNRS, 45071 Orléans cedex 2, France
        \and
        Station de Radioastronomie de Nançay, Observatoire de Paris, PSL Research University,
        CNRS, Université d’Orléans, OSUC, 18330 Nançay, France
        \label{nancay}
        \and
        SKA South Africa, The Park, Park Road, Pinelands 7405, South Africa
        \label{skasa}
        \and
        Department of Physics and Astronomy, University of the Western Cape,
        Private Bag X17, Bellville 7535, South Africa
        \label{uniwesterncape}
        \and
        Jodrell Bank Centre for Astrophysics, Department of Physics and Astronomy,
        The University of Manchester, Manchester M13 9PL, UK
        \label{manchester}
        \and
        Technische Universität Berlin, Institut für Geodäsie und
        Geoinformationstechnik, Fakultät VI, Sekr. H 12, Straße des 17. Juni
        135, 10623 Berlin, Germany
        \label{tuberlin}
        \and
        GFZ German Research Centre for Geosciences, Telegrafenberg, 14473 Potsdam, Germany
        \label{gfz}
        \and
        SKA Organisation, Jodrell Bank, Macclesfield SK11 9FT, UK
        \label{skajb}
        \and
        \label{lebedev}
        Astro Space Centre, Lebedev Physical Institute, Russian Academy of Sciences,
        Profsoyuznaya Str. 84/32, Moscow 117997, Russia
        \and
        CSIRO Astronomy and Space Science, PO Box 1130 Bentley, WA 6102, Australia
        \label{csiro}
        \and
        Canadian Institute for Theoretical Astrophysics, University of Toronto,
        60 St. George Street, Toronto, ON M5S 3H8, Canada
        \label{toronto}
        \and
        Anton Pannekoek Institute, University of Amsterdam, Postbus 94249,
        1090 GE Amsterdam, The Netherlands
        \label{uniamsterdam}
        \and
        \label{hamburg}
        Hamburger Sternwarte, University of Hamburg, Gojenbergsweg 112, 21029 Hamburg, Germany
        \and
        \label{garching}
        Max-Planck-Institut für Astrophysik, Karl-Schwarzschild-Straße 1, 85748 Garching b. München, Germany
        \and
        \label{tautenburg}
        Thüringer Landessternwarte, Sternwarte 5, 07778 Tautenburg, Germany
        \and
        \label{potsdam}
        Leibniz-Institut f\"ur Astrophysik Potsdam (AIP), An der Sternwarte 16, 14482 Potsdam, Germany
}
\date{Received 24 Sep 2020 / Accepted 15 Nov 2020}

\abstract
{
  Radio pulses from pulsars are affected by plasma dispersion,
  which results in a frequency-dependent propagation delay.
  Variations in the magnitude of this effect lead to
  an additional source of red noise in pulsar timing experiments,
  including \acp{pta} that aim to detect nanohertz gravitational waves.
}
{
  We aim to quantify the time-variable dispersion with much improved
  precision and characterise the spectrum of these variations.
}
{
  We use the pulsar timing technique
  to obtain highly precise \ac{dm} time series.
  Our dataset consists of observations of
  \npsr{} \acp{msp}, which were observed for
  up to \maxspan{} years with the \ac{lofar} telescope
  at a centre frequency of $\sim150$\,MHz.
  Seventeen of these sources were observed
  with a weekly cadence,
  while the rest were observed at monthly cadence.
}
{
  We achieve a median \ac{dm} precision
  of the order of $10^{-5}$\,\DMunits{}
  for a significant fraction of our sources.
  We detect significant variations of the \ac{dm}
  in all pulsars with a median \ac{dm} uncertainty
  of less than $2\times10^{-4}$\,\DMunits{}.
  The noise contribution to pulsar timing experiments
  at higher frequencies is calculated to be
  at a level of 0.1--\SI{10}{\micro\second} at 1.4\,GHz
  over a timespan of a few years,
  which is in many cases larger than the typical timing precision
  of \SI{1}{\micro\second} or better that \acp{pta} aim for.
  We found no evidence for a dependence of \ac{dm} on
  radio frequency for any of the sources in our sample.
}
{
  The \ac{dm} time series we obtained using \ac{lofar}
  could in principle be used to correct higher-frequency data for
  the variations of the dispersive delay.
  However, there is currently the practical restriction that pulsars tend to
  provide either highly precise \acp{toa} at 1.4\,GHz or
  a high \ac{dm} precision at low frequencies, but not both,
  due to spectral properties.
  Combining the higher-frequency \acp{toa} with those from \ac{lofar}
  to measure the infinite-frequency \ac{toa} and \ac{dm} would improve the result.
}

\maketitle
\acresetall

\section{Introduction}
Pulsars are highly magnetised, rapidly rotating neutron stars, the remnants of massive stars that ended their lives in a supernova explosion.
Pulsars emit beams of electromagnetic radiation, pronounced mostly at radio frequencies,
which sweep around in space as the neutron star rotates. 
If any of the beams cross the line of sight to the Earth, we detect regular pulses of radiation.
While the shape of individual pulses can vary significantly, the average pulse shape, integrated over hundreds of pulses, is usually stable, which allows for a precise measurement of the \acp{toa} of the integrated pulses.

There are two major distinct populations of pulsars: the canonical pulsars with pulse periods of the order of $\sim$0.1\,s to $\sim$20\,s and the \acp{msp}, with typical spin periods of 1.4\,ms to $\sim$30\,ms. The latter were `spun up' to shorter pulse periods by the transfer of mass and angular momentum from a binary stellar companion, which is why they are also called `recycled' pulsars \citep{tkb+15}. Millisecond pulsars are of particular scientific interest as the much shorter pulse periods allow for a more precise determination of the \acp{toa}, and their rotation was found to be much more stable than that of canonical pulsars \citep{hlk10,vbc+09}.

\paragraph{Pulsar timing.}
The high rotational stability of pulsars, and \acp{msp} in particular, together with the precise measurements afforded by pulsar timing, allows for extremely accurate modelling of the pulsar's astrometric and astrophysical properties with a so-called timing model, a method called pulsar timing \citep[see][]{lk05}. The model is usually evaluated by inspecting the timing residuals, that is, the difference between the measured \acp{toa} of the pulses and the ones predicted by the model.

\paragraph{Pulsar timing arrays.}
A major application of pulsars is in the so-called \ac{pta} projects, which aim to detect nanohertz frequency gravitational waves \citep[see, e.g.][]{hd17,tib18,btc+19}. The basic idea behind these experiments is that passing gravitational waves would distort the spacetime around Earth in a way that the arrival times of some pulsars would be delayed, while the pulses of other pulsars arrive earlier at the telescope. By observing a large number of spatially separated, precisely timed pulsars over a time span of years to decades, a gravitational wave would be measurable in the spatial correlation of the timing residuals of different pulsars. One major focus of the \acp{pta} is to detect a stochastic gravitational wave background signature such as the Hellings and Downs curve \citep{hd83}. This background is expected to be caused by inspiralling super-massive black hole binaries.
To achieve a high precision in these experiments, many sources of noise have to be taken into account, including propagation effects in the \ac{iism}.

\paragraph{Influence of the interstellar medium.}
The \ac{iism} is a cold plasma of ionised particles. An electromagnetic wave passing through this medium will experience a frequency-dependent change in group velocity, a phenomenon called \emph{dispersion}. Specifically, this induces an additional delay $\Delta t$ in the \acp{toa}, which can be approximated as \citep[see][]{lk05}:
\begin{equation}
	\label{eq:disp_delay}
	\Delta t = \frac{\rm DM}{k \nu^2},
\end{equation}
where $k \simeq 2.41\times 10^{-4}$\,cm$^{-3}$\,pc\,MHz$^{-2}$\,s$^{-1}$ is the dispersion constant\footnote{Although $k$ can be calculated more precisely, we use this rounded value as it is common practice in the field. See \cite{kul20} for a discussion.}, $\nu$ the observing frequency (expressed in MHz) and DM is the `dispersion measure' (the integrated electron column density along the line of sight, expressed in \DMunits{}), which is defined as:
\acused{dm}
\begin{equation}
	{\rm DM} = \int_0^d n_{\rm e}(l){\rm d}l,
\end{equation}
where $d$ is the distance to the pulsar and $n_{\rm e}$ is the electron density.

The \ac{iism} is inhomogeneous, and so as our line of sight to the pulsar changes due to its proper motion, the Earth's motion and the motion of the \ac{iism} we sample different regions of the \ac{iism},
which implies a temporal variability in the \ac{dm}.
The variations in the \ac{dm} lead to a variable dispersive delay (see Eq.~\ref{eq:disp_delay}) and, thus, are a major source of noise in pulsar timing.

Using relatively high observing frequencies (> 1.4\,GHz) minimises the impact of \ac{dm} variations on the \acp{toa} as the dispersive delay scales with $\nu^{-2}$ (see Eq.~\ref{eq:disp_delay}).
However, the \ac{iism} turbulence spectrum is steep with significantly more power at larger scales \citep{ars95}, so for long timing campaigns, the dispersive delays will sooner or later still have a significant impact on the \acp{toa}. Also, the observing frequency cannot be increased indefinitely, as pulsars have rather steep spectra 
\citep{mkkw20}, which leads to a loss of \ac{snr} at very high frequencies. Out of the 33 \acp{msp} in their sample, \cite{lkg+16} selected 12 candidates with promising spectral indices to run test observations at 5\,GHz (3 at 9\,GHz). All the selected sources are detected, but the \ac{snr} is significantly lower than at 1.4\,GHz, especially in the 9\,GHz band. However, some of the sources have a sub-\si{\micro\second} timing precision (post-fit RMS) at 5\,GHz, so this approach can work for some flat-spectrum sources.

A more generally applicable solution is to correct for the time-variable dispersive delays by subtracting them from the \acp{toa}, which effectively yields \acp{toa} at infinite frequency.
Dispersive delays can be measured either by using multi-frequency observations with several receivers, or by splitting wide-band observations into multiple sub-bands,
but this kind of observation is not always available in \ac{pta} experiments \citep[see, e.g.][]{dcl+16}.
Also, the correction for the \ac{dm} increases the uncertainty of the infinite-frequency \ac{toa} and this effect is especially large at high frequencies and small fractional bandwidths.

Another approach is to use low-frequency observations to measure the \ac{dm} very precisely and use these measurements to correct the higher-frequency \acp{toa}, which are often more sensitive and less affected by \ac{iism} effects.
One potential complication of this approach is the possibility of frequency-dependent \acp{dm} \citep{css16,dvt+19}, which are caused by the fact that due to interstellar scattering, 
low-frequency observations effectively sample a larger volume of the \ac{iism} and are sensitive to \ac{iism} structures on a larger scale than at high frequencies.
Therefore, dispersive delays measured at low frequency may not be representative for those experienced at high frequencies.
Additionally, time-variable scattering can cause variations in the shape of the pulse profile that, if not corrected for, cause a frequency-dependent delay in the timing. While scattering induces a delay scaling with $\nu^{-4}$, some of its signature can be absorbed into a \ac{dm} measurement, so this effect can cause a mismatch of the apparent \ac{dm} at different frequencies.

In this paper, we present precise \ac{dm} time series of \npsr{} \acp{msp} using low-frequency ($\sim$150\,MHz) observations with a particular focus on pulsars used in \ac{pta} experiments.
In Sect.~\ref{sec:obs} we describe our observational setup and our sources, while Sect.~\ref{sec:ana} explains our data analysis. We discuss our findings in Sect.~\ref{sec:disc} and present our conclusions in Sect.~\ref{sec:concl}.

\section{Observations}
\label{sec:obs}
We used observations taken with the \ac{ilt} \acp{hba}, described in detail by \cite{vwg+13}.
Specifically, we used the data from two different pulsar monitoring campaigns, both of which are still ongoing.
Campaign~1 uses the \ac{lofar} Core stations situated in the north-east of the Netherlands to observe the sources of interest with monthly cadence for up to \maxcorespan{} years between \firstcoreobsdate{} and \lastcoreobsdate{}. These data were coherently dedispersed and reduced using \ac{pulp} described by \cite{sha+11} and \cite{kvh+16}.
It produces data cubes with resolution in frequency (195.3125\,kHz-wide channels), time (10-sec sub-integrations), polarisation (four coherency products) and rotational phase (256 to 1024 phase bins).
The data are stored in \textsc{timer} format, which is similar to the \textsc{psrfits} format described in \cite{hvm04}.
Each observation uses a centre frequency of 148.9\,MHz, a bandwidth of 78.1\,MHz (400 frequency channels) and lasts 5 to 30 minutes, depending on the brightness of the pulsars.

In Campaign~2, the six \ac{lofar} stations of the \ac{glow} consortium, located in Effelsberg (telescope identifier DE601), Unterweilenbach (DE602), Tautenburg (DE603), Potsdam-Bornim (DE604), Jülich (DE605) and Norderstedt (DE609), were used as individual stand-alone telescopes, not connected to the ILT network. The beamformed data from the stations were sent to the \ac{mpifr} and the Forschungszentrum Jülich on dedicated high-speed links, where recording computers ran the dedicated \ac{lump}\footnote{Publicly available at
  \url{https://github.com/AHorneffer/lump-lofar-und-mpifr-pulsare} and described on
  \url{https://deki.mpifr-bonn.mpg.de/Cooperations/LOFAR/Software/LuMP}.}
data-taking software. \ac{lump} formats and otherwise prepares the beamformed pulsar data for subsequent (off-line but near-real-time) phase-resolved averaging (commonly referred to as `folding') using the \textsc{dspsr} software package \citep{vb11}. This produces data cubes in the same format as \ac{pulp}.
The data we used from this campaign were taken between \firstglowobsdate{} and \lastglowobsdate{}.
The resulting dataset of these observations covers a time span of up to \maxglowspan{} years per pulsar, with a weekly cadence and typical integration times of 1 to 3 hours.
Due to the difference in collecting areas (a factor of $\sim$10), the faintest pulsars were only observed with the Core and are not detectable with the international stations.
Early observations and the observations of DE601 have a total bandwidth of 95.3\,MHz (488 frequency channels), centred at 149.9\,MHz. For technical reasons, the bandwidth of the other stations was reduced to 71.5\,MHz (366 channels) in February 2015. In order to minimise the impact of the bandwidth reduction on the scientific quality of our data, the centre frequency was shifted to align the observed bandwidth with the most sensitive part of the bandpass, resulting in a new centre frequency of 153.8\,MHz. This implies a shift in centre frequency by an integer number of frequency channels (20), so that the frequencies of individual channels remained constant over the entire dataset.

Table~\ref{obs_table} shows detailed information on the observation characteristics for each source.
We excluded PSR~J1939+2134 (PSR~B1937+21) from our analysis because its strongly variable scattering has a significant impact on the \ac{dm} estimation \citep[e.g.\ it shows the largest variation in scattering time of all sources in][]{lmj+16}. Also, due to the very strong scattering, the scattering tail of the pulse profile merges with the next pulse, so the profile is very wide \citep[see Fig.~2 of][]{kvh+16}. Observations of this pulsar at higher frequencies or more advanced analysis techniques like cyclic spectroscopy \citep{dem11} could lead to more robust and similarly precise \ac{dm} measurements.

\begin{table*}
  \centering
  \caption{Summary of observations.
    Given are the pulsar name in J2000 coordinates,
    the pulse period $P$, the catalogue \ac{dm},
    the ecliptic latitude $\beta$,
    the total time span of the observations,
    the number of observations with \ac{glow} and the \ac{lofar} Core
    $N_\text{glow}$ and $N_\text{core}$,
    the median frequency-integrated \ac{toa} uncertainty,
    the median uncertainty of individual \ac{dm} measurements,
    the median reduced $\chi^2$ of the individual \ac{dm} fits,
    and the number of frequency channels in each observation.
    The last column shows whether the pulsar is used
    in different \ac{pta} projects
    \citep[i.e.\ 
      E: \acl{epta}, \acs{epta};
      P: \acl{ppta}, \acs{ppta};
      N: \acl{nanograv}, \acs{nanograv};
      see][]
          {vlh+16,abb+18}.
    \acused{epta}\acused{ppta}\acused{nanograv}
    A pulsar observed by any of the \acp{pta} is also used by the \acf*{ipta}.
  }
  \label{obs_table}
  \begin{tabular}{ c c c c c c c c c c c c}
source~name   &  $P$          &  DM                                            &  $\beta$       &  span        &  $N_\text{glow}$  &  $N_\text{core}$  &  med($\sigma_\text{ToA}$)  &  med($\sigma_\text{DM}$)                              &  med$\left(\frac{\chi^2}{n_\text{free}}\right)$  &  $N_\text{chan}$  &  PTA                        \\
(J2000)       &  (ms)         &  $\left(\frac{\text{pc}}{\text{cm}^3}\right)$  &  (deg)         &  (yrs)       &  ~                &  ~                &  (\si{\micro\second})      &  $\left(10^{-5}\frac{\text{pc}}{\text{cm}^3}\right)$  &  ~                                               &  ~                &  ~                          \vspace{4pt}  \\  \hline  \rule{0pt}{15pt}\unskip
J0030+0451    &  \dcol{4.9}   &  \dcol{4.3}                                    &  \dcol{1.4}    &  \dcol{7.0}  &  355              &  79               &  7                         &  16                                                   &  \dcol{1.0}                                      &  10               &  E~\phantom{P}~N            \\
J0034$-$0534  &  \dcol{1.9}   &  \dcol{13.8}                                   &  \dcol{-8.5}   &  \dcol{7.0}  &  1041             &  78               &  1                         &  3                                                    &  \dcol{1.0}                                      &  20               &  E~\phantom{P}~\phantom{N}  \\
J0218+4232    &  \dcol{2.3}   &  \dcol{61.2}                                   &  \dcol{27.0}   &  \dcol{6.5}  &  539              &  74               &  13                        &  35                                                   &  \dcol{0.9}                                      &  10               &  E~\phantom{P}~\phantom{N}  \\
J0407+1607    &  \dcol{25.7}  &  \dcol{35.6}                                   &  \dcol{-4.7}   &  \dcol{6.4}  &  386              &  65               &  24                        &  56                                                   &  \dcol{1.0}                                      &  10               &  -                          \\
J0621+1002    &  \dcol{28.9}  &  \dcol{36.5}                                   &  \dcol{-13.3}  &  \dcol{7.0}  &  10               &  74               &  251                       &  568                                                  &  \dcol{1.3}                                      &  10*              &  E~\phantom{P}~\phantom{N}  \\
J0645+5158    &  \dcol{8.9}   &  \dcol{18.2}                                   &  \dcol{28.9}   &  \dcol{6.9}  &  268              &  4                &  3                         &  7                                                    &  \dcol{0.9}                                      &  10               &  \phantom{E}~\phantom{P}~N  \\
J0740+6620    &  \dcol{2.9}   &  \dcol{15.0}                                   &  \dcol{44.1}   &  \dcol{4.8}  &  261              &  0                &  2                         &  4                                                    &  \dcol{0.9}                                      &  10               &  \phantom{E}~\phantom{P}~N  \\
J0751+1807    &  \dcol{3.5}   &  \dcol{30.2}                                   &  \dcol{-2.8}   &  \dcol{7.1}  &  0                &  71               &  34                        &  89                                                   &  \dcol{0.6}                                      &  10*              &  E~\phantom{P}~\phantom{N}  \\
J0952$-$0607  &  \dcol{1.4}   &  \dcol{22.4}                                   &  \dcol{-17.9}  &  \dcol{2.9}  &  37               &  33               &  9                         &  14                                                   &  \dcol{1.0}                                      &  8*               &  -                          \\
J1012+5307    &  \dcol{5.3}   &  \dcol{9.0}                                    &  \dcol{38.8}   &  \dcol{7.1}  &  1210             &  79               &  4                         &  12                                                   &  \dcol{1.2}                                      &  10               &  E~\phantom{P}~N            \\
J1022+1001    &  \dcol{16.5}  &  \dcol{10.3}                                   &  \dcol{-0.1}   &  \dcol{7.1}  &  1224             &  80               &  5                         &  12                                                   &  \dcol{1.2}                                      &  20               &  E~P~N                      \\
J1024$-$0719  &  \dcol{5.2}   &  \dcol{6.5}                                    &  \dcol{-16.0}  &  \dcol{7.1}  &  5                &  81               &  32                        &  80                                                   &  \dcol{0.9}                                      &  8*               &  E~P~N                      \\
J1125+7819    &  \dcol{4.2}   &  \dcol{11.2}                                   &  \dcol{62.5}   &  \dcol{4.8}  &  282              &  0                &  3                         &  8                                                    &  \dcol{0.9}                                      &  10               &  \phantom{E}~\phantom{P}~N  \\
\phantom{$^a$}J1300+1240$^a$    &  \dcol{6.2}   &  \dcol{10.2}                                   &  \dcol{17.6}   &  \dcol{7.1}  &  377              &  80               &  3                         &  7                                                    &  \dcol{0.8}                                      &  20               &  -                          \\
J1400$-$1431  &  \dcol{3.1}   &  \dcol{4.9}                                    &  \dcol{-2.2}   &  \dcol{4.3}  &  145              &  49               &  7                         &  6                                                    &  \dcol{0.8}                                      &  8*               &  -                          \\
J1544+4937    &  \dcol{2.2}   &  \dcol{23.2}                                   &  \dcol{65.9}   &  \dcol{5.6}  &  5                &  53               &  19                        &  59                                                   &  \dcol{1.0}                                      &  6*               &  -                          \\
J1552+5437    &  \dcol{2.4}   &  \dcol{22.9}                                   &  \dcol{70.7}   &  \dcol{3.7}  &  0                &  41               &  3                         &  7                                                    &  \dcol{0.9}                                      &  10*              &  -                          \\
J1640+2224    &  \dcol{3.2}   &  \dcol{18.4}                                   &  \dcol{44.1}   &  \dcol{7.1}  &  435              &  79               &  7                         &  18                                                   &  \dcol{0.8}                                      &  8                &  E~\phantom{P}~N            \\
J1658+3630    &  \dcol{33.0}  &  \dcol{3.0}                                    &  \dcol{58.7}   &  \dcol{2.6}  &  164              &  6                &  7                         &  18                                                   &  \dcol{0.9}                                      &  10               &  -                          \\
J1713+0747    &  \dcol{4.6}   &  \dcol{16.0}                                   &  \dcol{30.7}   &  \dcol{7.0}  &  7                &  79               &  12                        &  29                                                   &  \dcol{0.6}                                      &  10*              &  E~P~N                      \\
J1730$-$2304  &  \dcol{8.1}   &  \dcol{9.6}                                    &  \dcol{0.2}    &  \dcol{6.5}  &  0                &  74               &  21                        &  49                                                   &  \dcol{0.8}                                      &  10*              &  E~P~\phantom{N}            \\
J1738+0333    &  \dcol{5.9}   &  \dcol{33.8}                                   &  \dcol{26.9}   &  \dcol{7.1}  &  0                &  77               &  39                        &  110                                                  &  \dcol{0.7}                                      &  6*               &  E~\phantom{P}~N            \\
J1744$-$1134  &  \dcol{4.1}   &  \dcol{3.1}                                    &  \dcol{11.8}   &  \dcol{7.0}  &  372              &  79               &  14                        &  19                                                   &  \dcol{0.9}                                      &  10               &  E~P~N                      \\
J1853+1303    &  \dcol{4.1}   &  \dcol{30.6}                                   &  \dcol{35.7}   &  \dcol{6.5}  &  1                &  72               &  21                        &  50                                                   &  \dcol{0.8}                                      &  8*               &  E~\phantom{P}~N            \\
\phantom{$^b$}J1857+0943$^b$    &  \dcol{5.4}   &  \dcol{13.3}                                   &  \dcol{32.3}   &  \dcol{7.1}  &  0                &  75               &  41                        &  82                                                   &  \dcol{1.3}                                      &  10*              &  E~P~N                      \\
J1911$-$1114  &  \dcol{3.6}   &  \dcol{31.0}                                   &  \dcol{11.1}   &  \dcol{7.0}  &  0                &  74               &  32                        &  68                                                   &  \dcol{0.9}                                      &  10               &  E~\phantom{P}~\phantom{N}  \\
J1918$-$0642  &  \dcol{7.6}   &  \dcol{26.6}                                   &  \dcol{15.4}   &  \dcol{7.0}  &  0                &  75               &  55                        &  136                                                  &  \dcol{0.8}                                      &  10*              &  E~\phantom{P}~N            \\
J1923+2515    &  \dcol{3.8}   &  \dcol{18.9}                                   &  \dcol{46.7}   &  \dcol{6.6}  &  0                &  73               &  6                         &  15                                                   &  \dcol{0.9}                                      &  10*              &  \phantom{E}~\phantom{P}~N  \\
J1944+0907    &  \dcol{5.2}   &  \dcol{24.4}                                   &  \dcol{29.9}   &  \dcol{6.5}  &  35               &  64               &  33                        &  80                                                   &  \dcol{0.8}                                      &  6*               &  \phantom{E}~\phantom{P}~N  \\
\phantom{$^c$}J1955+2908$^c$    &  \dcol{6.1}   &  \dcol{104.5}                                  &  \dcol{48.7}   &  \dcol{6.6}  &  0                &  74               &  100                       &  273                                                  &  \dcol{0.9}                                      &  10*              &  E~\phantom{P}~N            \\
J2043+1711    &  \dcol{2.4}   &  \dcol{20.7}                                   &  \dcol{34.0}   &  \dcol{6.7}  &  3                &  71               &  6                         &  17                                                   &  \dcol{0.9}                                      &  10*              &  \phantom{E}~\phantom{P}~N  \\
J2051$-$0827  &  \dcol{4.5}   &  \dcol{20.7}                                   &  \dcol{8.8}    &  \dcol{6.5}  &  6                &  74               &  14                        &  29                                                   &  \dcol{0.7}                                      &  10*              &  -                          \\
J2145$-$0750  &  \dcol{16.1}  &  \dcol{9.0}                                    &  \dcol{5.3}    &  \dcol{7.0}  &  1010             &  81               &  3                         &  9                                                    &  \dcol{1.1}                                      &  20               &  E~P~N                      \\
J2222$-$0137  &  \dcol{32.8}  &  \dcol{3.3}                                    &  \dcol{8.0}    &  \dcol{3.8}  &  130              &  41               &  22                        &  49                                                   &  \dcol{0.9}                                      &  10*              &  -                          \\
J2302+4442    &  \dcol{5.2}   &  \dcol{13.7}                                   &  \dcol{45.7}   &  \dcol{6.2}  &  0                &  71               &  20                        &  51                                                   &  \dcol{0.8}                                      &  10*              &  \phantom{E}~\phantom{P}~N  \\
J2317+1439    &  \dcol{3.4}   &  \dcol{21.9}                                   &  \dcol{17.7}   &  \dcol{7.0}  &  381              &  77               &  2                         &  6                                                    &  \dcol{1.0}                                      &  10               &  E~\phantom{P}~N            \\
\hline
\end{tabular}
 \\
  \vspace{6pt}
  \raggedright
  \hspace{15pt} * For these pulsars, an analytic standard template
  without frequency resolution was used. \\
  \hspace{15pt} $^a$ PSR~B1257+12 \\
  \hspace{15pt} $^b$ PSR~B1855+09 \\
  \hspace{15pt} $^c$ PSR~B1953+29
\end{table*}

\section{Data analysis}

\label{sec:ana}
\subsection{Pre-processing}
The basic processing in this work has been carried out with the \textsc{psrchive} \citep{hvm04,vdo12} software package.
As a first step, the data were cleaned from \ac{rfi}, by using a modified version of the `surgical' algorithm of the \textsc{clean.py} script from the \textsc{coastguard} \citep{lkg+16} python package\footnote{The version we used was provided by \cite{kun17} and is publicly available at \url{https://github.com/larskuenkel/iterative_cleaner}.} to remove affected frequency channels and sub-integrations. In the rare case of outliers due to remaining \ac{rfi}, the observations were also manually inspected and additional cleaning was applied using the \textsc{psrchive} program \textsc{pazi}. On average, \meanrfifrac{}\% of data were removed in this process.

Pulse profiles can be polarisation dependent, so the total intensity profile can be significantly distorted if the different polarisation channels are not calibrated correctly, which affects the \acp{toa}. As \ac{lofar} antennas are static, there are several time-dependent projection effects to take into account.
The data were calibrated in polarisation following the methods outlined in \cite{nsk+15} using the \textsc{dreambeam}\footnote{Publicly available at \url{https://github.com/2baOrNot2ba/dreamBeam}, by T.~Carozzi.} python package to calculate the Jones matrices. For some pulsars (especially PSRs~J1022+1001 and J2145$-$0750) this significantly improved the reduced $\chi^2$ values of the \ac{dm} fits.

To make the dataset as homogeneous as possible, the bandwidth of all observations has been reduced to 70.3\,MHz (i.e.\ 360 channels, centred at 153.2\,MHz). To achieve this, two empty dummy channels had to be added at the top of the band for the \ac{lofar} Core observations.

\subsection{Timing}
For the 20 pulsars also analysed by the \ac{ipta}, we took the timing model from the first \ac{ipta} data release \citep[][data combination B]{vlh+16}\footnote{Publicly available at \url{http://www.ipta4gw.org/}} and removed any existing \ac{dm} models (including Solar wind) or FD parameters \citep[which describe a frequency-dependent, non-dispersive delay in the timing residuals, see][]{abb+15}, as we are interested in the time evolution of the \ac{dm} and using frequency-resolved timing removes the need for FD parameters. To account for the different time span over which the \ac{ipta} ephemerides were derived, and because the \ac{dm} estimates are expected to be more precise at the low radio frequencies of \ac{lofar}, we performed an initial timing analysis over our dataset, with the aim of updating the reference \ac{dm} for each pulsar.
For this initial timing analysis, we derived an analytic template from a single, bright observation by fitting a series of von Mises functions \citep[see, e.g.][]{js01} to the total intensity profile of the observation using the program \textsc{paas}. The functions are implemented in the form:
\begin{equation}
  f(x) = A \cdot e^{\kappa \cdot (\cos (2\pi (x-\mu)) -1)},
\end{equation}
with $A$ being the amplitude of the component, $\kappa$ the compactness and $\mu$ the pulse phase. The phase $x$ is defined such that one full rotation corresponds to $x = 1$.
To obtain the \acp{toa}, we used the \textsc{fdm}\footnote{This algorithm is identical to that described by \cite{tay92}, except for the uncertainties. \textsc{fdm} uses either formal uncertainties or Monte-Carlo simulations. We used the uncertainties determined from Monte-Carlo simulations.} algorithm as implemented in the program \textsc{pat} on copies of our observations with reduced frequency resolution (integrated down to 10 frequency channels).
Using these initial \acp{toa}, we calculated the \ac{dm} and its running average for each observation as described in Sect.~\ref{sec:calcDM}, to be able to phase-align observations affected by \ac{dm} variations.

For the 16 pulsars without a published \ac{ipta} timing model, we took our initial timing model from the ATNF Pulsar Catalogue (\textsc{psrcat})\footnote{\url{http://www.atnf.csiro.au/people/
  pulsar/psrcat/}} by \cite{mhth05}. For PSR~J1658+3630, there was no timing model available in the catalogue, so we used the timing model from \cite{scb+19}.
As these timing models often did not phase-align our dataset, we used the \tempoII{} software \citep{hem06} to fit for parameters describing the pulsar's rotation and, if needed, the position of the pulsar and its orbit in binary systems.


Using the optimised timing model and initial \ac{dm} time series, we phase-aligned our dataset and created a standard pulse profile template for each pulsar by averaging only the observations from the observing site with the most observations of that pulsar.
As the \ac{snr} of most observations is rather low (rarely above 100), we integrated the template in frequency to between \nchanmin{} and \nchanmax{} channels, depending on the pulsar's \ac{snr}. We then applied a wavelet smoothing algorithm to the template with the program \textsc{psrsmooth} \citep{dfg+13}, to avoid self-standarding \citep[see Appendix~A of][]{hbo05a}.
With this template, we again used the \textsc{fdm} algorithm to calculate the \acp{toa}, matching each frequency channel of the observation (that was frequency-integrated accordingly) with the corresponding channel of the template. With this method, any constant frequency dependence of the pulse profile is modelled by the template and will not affect our measurement of \ac{dm}.

If the templates generated in the procedure above were too noisy (due to low intrinsic flux of the pulsar and a small number of observations), the smoothing produced unphysical features in the profile shape. In these cases we used an analytic template without frequency resolution, created in the same way as in the initial analysis. In principle, this would imply the need of FD parameters in the timing model to account for the frequency-dependent profile shape, but in practice the frequency dependence is not significant, demonstrated by the median $\chi^2$ of the \ac{dm} fits being close to unity (see Table~\ref{obs_table}).

\subsection{Calculation of the DM}
\label{sec:calcDM}
To get a precise time series of the \ac{dm}, we used \tempoII{} to fit for \ac{dm} for each observation individually. This approach avoids any correlation of the measured \acp{dm} with other (time-dependent) timing parameters.
To mitigate the impact of outlier \acp{toa} on our measured \ac{dm}, we apply an automatic \ac{toa} rejection scheme as was done similarly in \cite{tvs+19}.
We reject a \ac{toa} if its residual-to-uncertainty ratio $k$ is larger than four times the mean $k$.
Specifically, we repeatedly applied the following rules until no \ac{toa} was removed in the iteration:

We calculated the timing residuals $r_i$ for each channel $i$ of a given observation using the \tempoII{} \textsc{general}\oldstylenums{2} plugin and noted the corresponding \ac{toa} uncertainties $\sigma_i$.
The mean residual-to-uncertainty ratio is then calculated as $k = \text{mean}( |r_i| / \sigma_i )$ and set to unity if the result was less than unity.
If the condition $|r_i| / \sigma_i > 4k$ was met for any of the residuals, we rejected the corresponding \ac{toa}.
Finally, we fit for \ac{dm} with \tempoII{}.

The iterative process is necessary as the variable \ac{dm} gives rise to significant structure in our frequency-resolved residuals, meaning that many are far from zero.
This is also why the factor $k$ was introduced, as without it, \acp{toa} at the edges of the band would be removed in the presence of a dispersive slope.
The outliers were caused by a low \ac{snr} in the corresponding frequency channel, which can occur due to the pulsar being intrinsically faint at that frequency, short integration times, removal of large parts of that frequency range due to \ac{rfi}, remaining low-level \ac{rfi} or a combination of these. On average, \meanprej\% of the \acp{toa} of an observation were removed during this step, mostly at the edges of the band where the telescope is less sensitive.
We estimate that our outlier rejection procedure will produce fewer than ten false positives across our entire \npsr{}-pulsar dataset.
While the median reduced $\chi^2$ of the final \ac{dm} fits is close to unity for all pulsars in our sample (see Table~\ref{obs_table}) and the distribution is strongly peaked around 1, there are observations not following this distribution with reduced $\chi^2 \gg 1$. These cases occured when most \acp{toa} of an observation were flawed, usually because the pulsar was too faint or the entire observation was dominated by \ac{rfi}. We excluded all \ac{dm} measurements with a reduced $\chi^2 > 5$ from our subsequent analysis (on average 4\% of all observations).
Finally, we applied the standard \tempoII{} procedure of multiplying the \ac{dm} uncertainty by the square root of the reduced $\chi^2$, because a high reduced $\chi^2$ indicates unmodelled structure or underestimated \ac{toa} uncertainties. To be conservative in our uncertainties and due to the low number of \acp{toa} per fit, we only applied this procedure for fits with reduced $\chi^2 > 1$.

\begin{figure*}
  \centering
  \includegraphics[width=\textwidth]{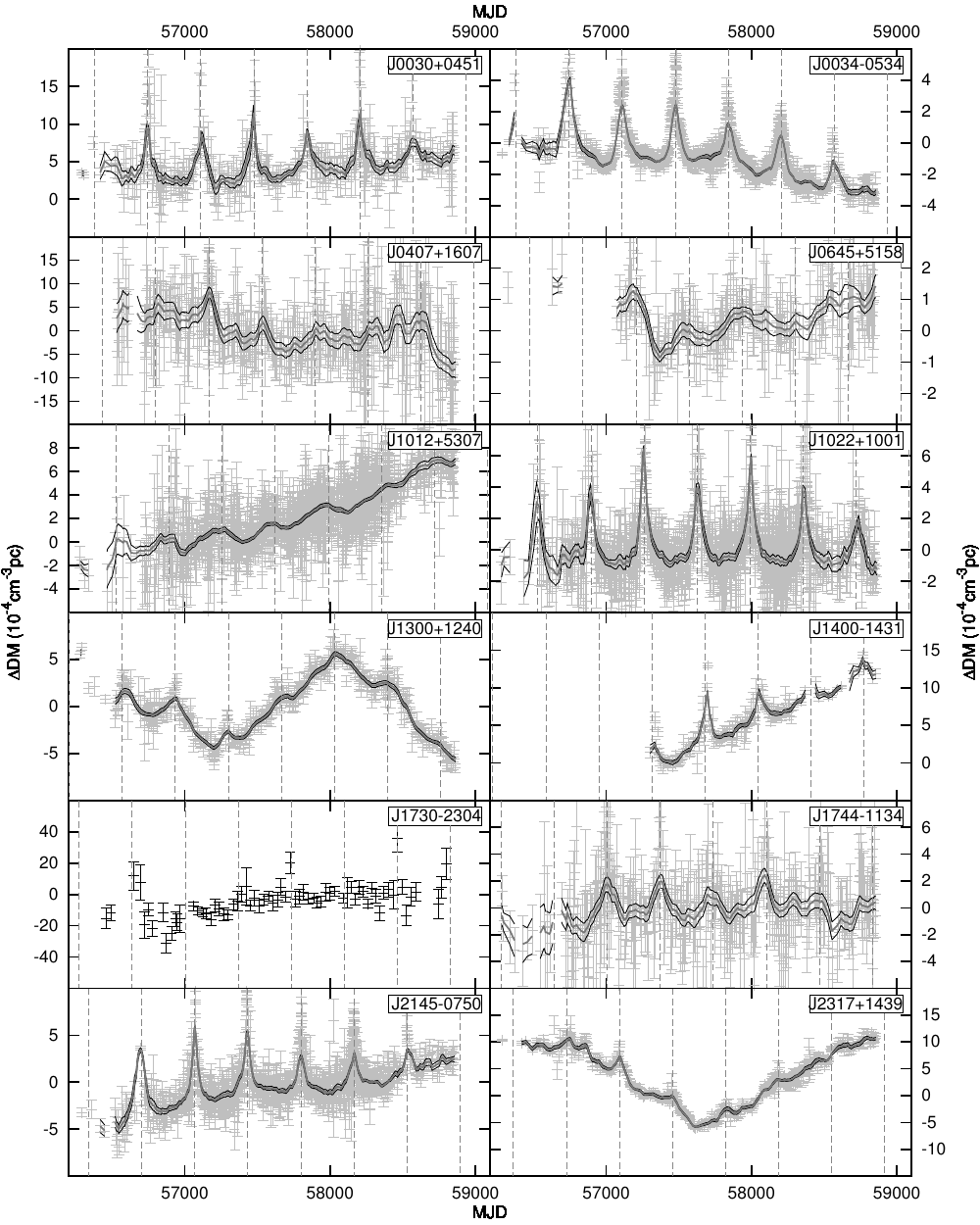}
  \caption{\ac{dm} time series for \npsrsolart{} \acp{msp}
    that show a detectable Solar wind signal.
    Individual \ac{dm} measurements are plotted in grey,
    while running averages and their uncertainty 
    are represented by solid grey and black lines, respectively.
    For faint pulsars only observed with the \ac{lofar} core,
    the running average is not plotted due to the low observing cadence.
    Only points with uncertainties less than three times
    the median uncertainty are plotted.
    The vertical dashed grey lines indicate the epochs
    of minimum angular separation from the Sun.
  }
  \label{fig:DMvars_solar}
\end{figure*}

\begin{figure*}
  \centering
  \includegraphics[width=\textwidth]{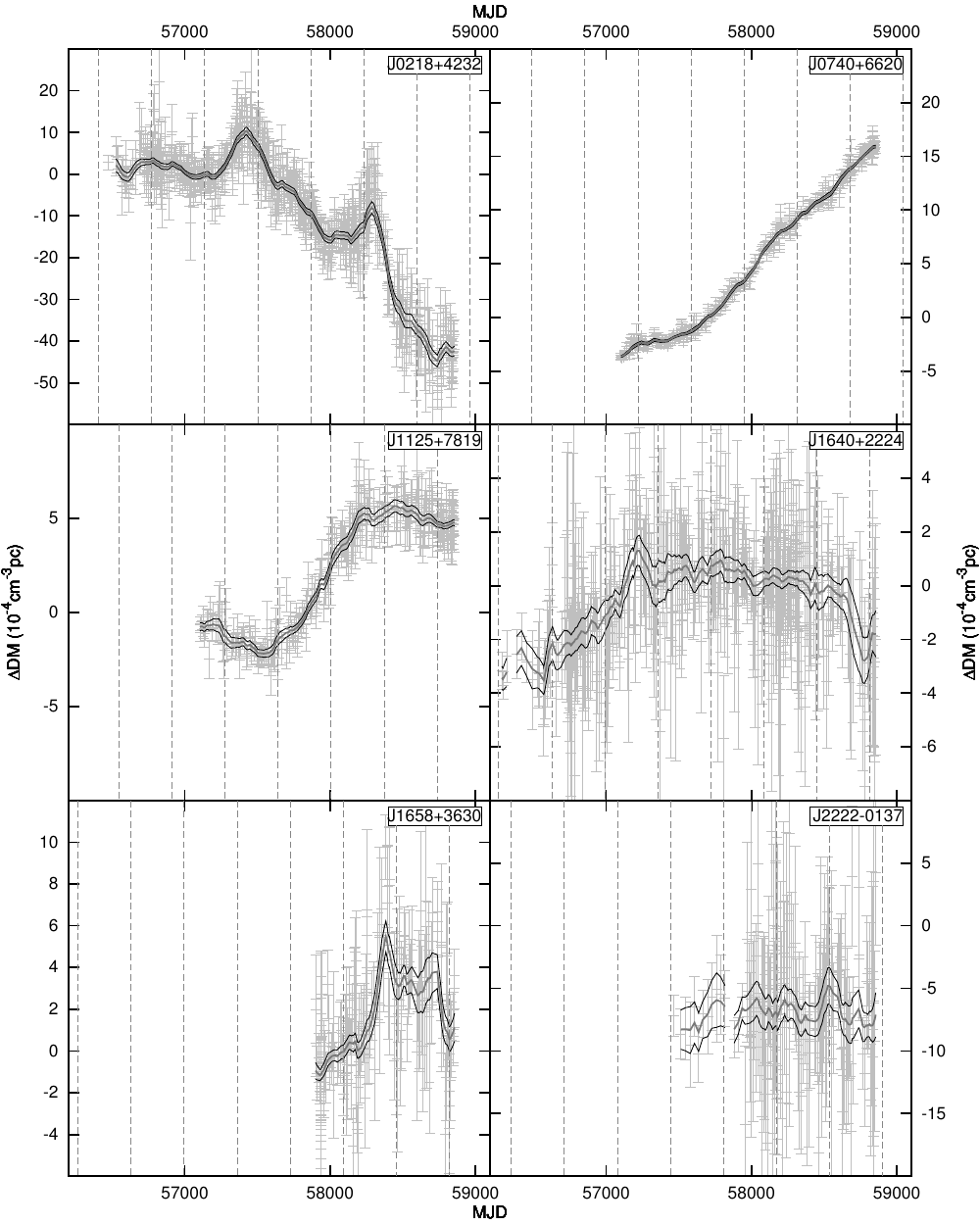}
  \caption{\ac{dm} time series for \npsrglowt{} \acp{msp}
    that were regularly observed with \ac{glow}
    and do not show a clear Solar wind signal.
    The representation is the same as in Fig.~\ref{fig:DMvars_solar}.
  }
  \label{fig:DMvars_glow}
\end{figure*}

\begin{figure*}
  \centering
  \includegraphics[width=\textwidth]{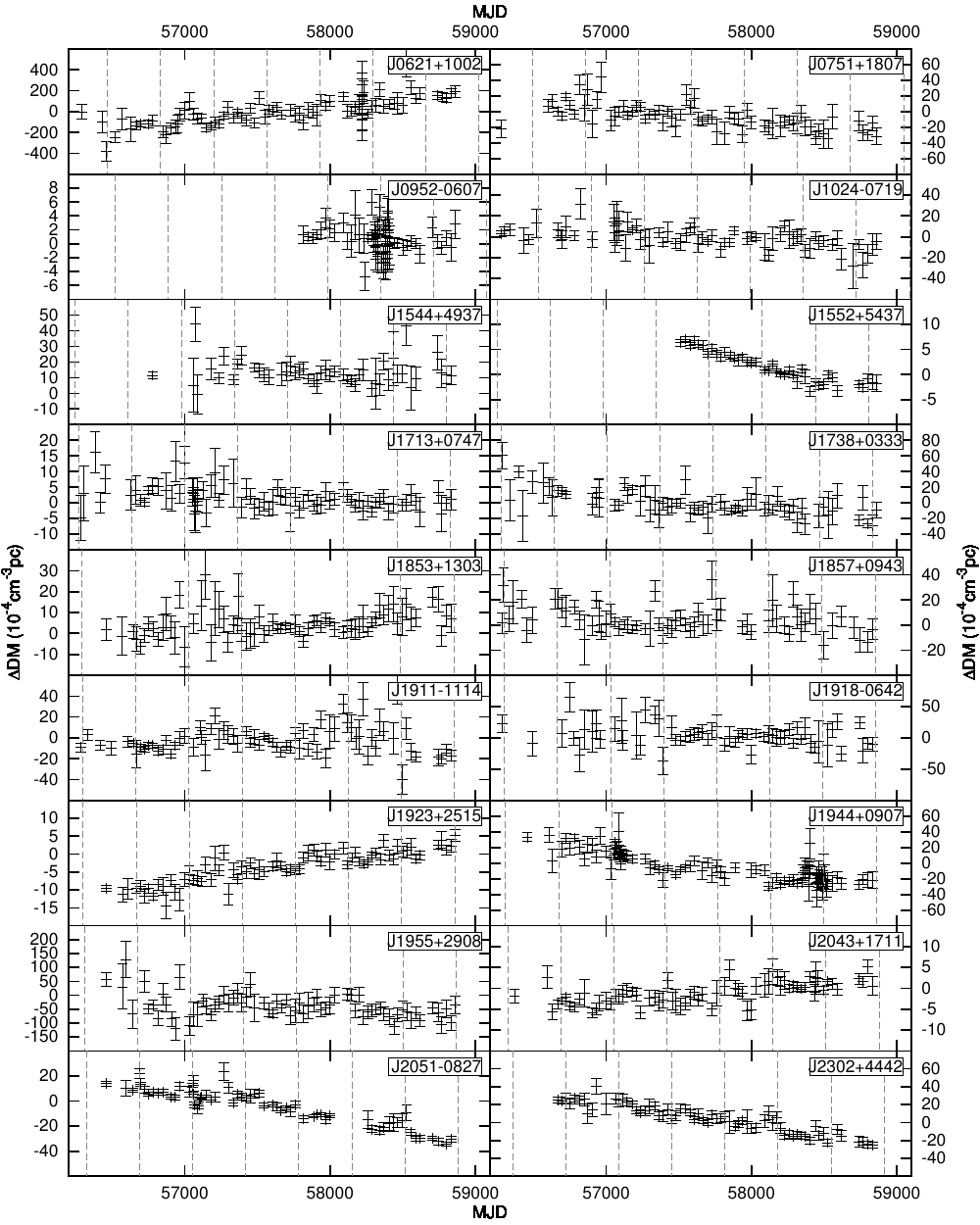}
  \caption{\ac{dm} time series for \npsrcoret{} faint \acp{msp}
    that are only monitored regularly with the \ac{lofar} Core
    and do not show a clear Solar wind signal.
    The representation is the same as in Fig.~\ref{fig:DMvars_solar}.
  }
  \label{fig:DMvars_core}
\end{figure*}

The resulting \ac{dm} time series are shown in Figs.~\ref{fig:DMvars_solar}-\ref{fig:DMvars_core}, with the reference \ac{dm} of the standard template subtracted. The median \ac{dm} precision for each source is shown in Table~\ref{obs_table}.
As the median reduced $\chi^2$ is close to unity for all pulsars in our sample, the data are well fit by the model. From that we can rule out any significant frequency-dependent structure in the residuals or a frequency dependence of the \ac{dm} as discussed in \cite{dvt+19}, where we present a system with more extreme \ac{dm} variations.

To improve our sensitivity to low-amplitude variations, we computed a running average of the \ac{dm} time series. For each observation MJD, a weighted average over all observations was formed, weighting each \ac{dm} value by the inverse of its variance. Additionally, the weights were reduced exponentially with time, scaled to $1/e$ of their original value over half the averaging window. We chose an averaging window of 30\,days for pulsars with sharp features in their \ac{dm} signal (i.e.\ PSRs~\ravgwindowlowlist{}) and 60\,days for the rest.

\section{Discussion}
\label{sec:disc}
Most of the pulsars in our sample show \ac{dm} variability, on diverse timescales. The amplitude of these variations ranges from $\sim10^{-4}$\,\DMunits{} to $10^{-3}$\,\DMunits{} over several years or less. Pulsars that do not show \ac{dm} variations (e.g.\ PSR~J1853+1303) are very faint at \ac{lofar} frequencies, with a single-observation \ac{dm} uncertainty of $2\times10^{-4}$\,\DMunits{} or worse, so the non-detection of variations may be caused by a lack of sensitivity, and variations of the order of a few $10^{-4}$\,\DMunits{} cannot be ruled out. 

\subsection{Solar wind}
We detect the \ac{dm} variations due to the Solar wind in \npsrsolart{} pulsars,
namely PSRs~\solarpsrlisttext{} (see Fig.~\ref{fig:DMvars_solar}).
The angular distance of these sources from the ecliptic ranges from \solarpsrlowestelatv{}\degree{} (PSR~\solarpsrlowestelat{}) to \solarpsrhighestelatv{}\degree{} (PSR~\solarpsrhighestelat{}, see Table~\ref{obs_table}). We note that for PSR~J1730$-$2304, there are only a few observations close to the Sun due to the low observing cadence with the \ac{lofar} Core and thus those observations clearly stand out from the rest of the \ac{dm} measurements.
For some pulsars with low (<$10\degree$) absolute ecliptic latitude, we do not clearly see the impact of the Solar wind due to a lack of measurement precision:
PSRs~\solarnondetectionlisttext{}.

As shown in \cite{tvs+19}, the widely used Solar wind models are insufficient to correctly account for the highly variable Solar wind, which leads to short-term variations in the residual \acp{dm} of the order of a few $10^{-4}$\,\DMunits{}, even at larger separations from the Sun (up to $\sim50\degree$).
From our study, we can confirm that a spherical model with constant amplitude (as is usually applied) does not suffice in most cases. This becomes clear in Fig.~\ref{fig:DMvars_solar}, where the amplitude of the Solar wind strongly varies from year to year, which is expected due to the variable Solar activity. Further analysis of the Solar wind impact on these data will be presented by Tiburzi et al. (in prep).

\subsection{Observed DM variations}
\label{sec:dm_features}
\cite{leg+18} report a \ac{dm} event in PSR~J1713+0747 around MJD~57510, where the \ac{dm} temporarily drops by about $4\times10^{-4}$\,\DMunits{}. There is no indication of this event in our \ac{lofar} \ac{dm} time series (see Fig.~\ref{fig:DMvars_core}). This can partly be explained by a lack of sensitivity as our \ac{dm} precision is similar to the amplitude of the event. Additionally, the event is observed at higher frequencies and could be partly smeared at lower frequencies due to a frequency-dependent \ac{dm}, as discussed in \cite{css16}.

Some sources show very linear gradients in \ac{dm}, for example PSRs~J0740+6620, J1552+5437, and J1400$-$1431, the last of which is also affected by the Solar wind. While these linear trends are unlikely to frequently occur solely due to a turbulent process, they can be explained by the pulsar's motion along the line of sight \citep{lcc+16}.
To quantify overall \ac{dm} trends, we used a least-squares fitting routine to fit a linear gradient to the \ac{dm} time series.
For the sources affected by the Solar wind, we first subtracted a spherically symmetric Solar wind model with variable amplitude, which was fit year by year to the difference between the measured \acp{dm} and a cubic spline derived from measurements furthest from the Sun.
The resulting \ac{dm} gradients are given in Table~\ref{tableII}.

\begin{figure*}
  \centering
  \includegraphics[width=\textwidth]{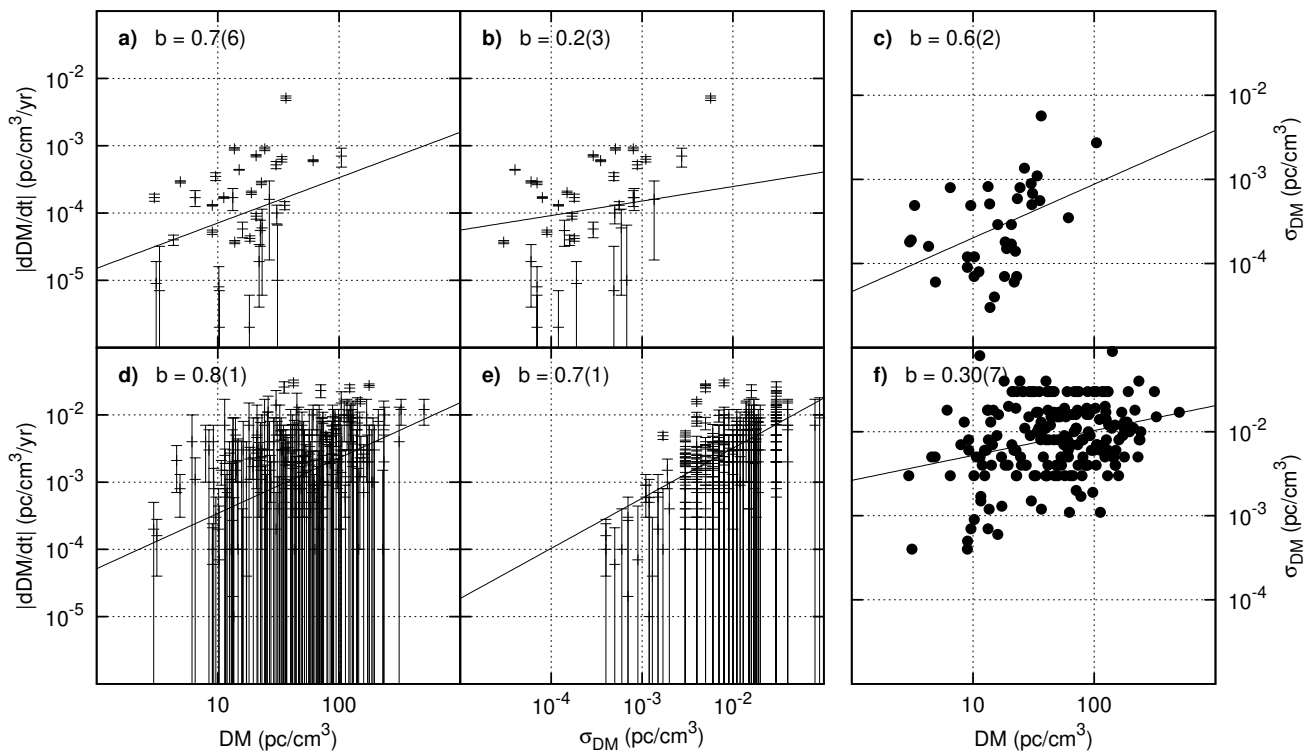}
  \caption{Correlations between DM, |dDM/d$t$| and $\sigma_\text{DM}$
    in our dataset (panels a to c) and that of \cite{hlk+04} (panels d to f).
    For each panel, we fit a power law of the form $y = a x^b$ to the data
    and the slope $b$ of this fit is given in each panel, along with its uncertainty.
    For the correlation between DM and $\sigma_\text{DM}$, we fit in logspace
    due to the lack of an uncertainty in $y$.
    As it was done by \cite{hlk+04}, we excluded all pulsars from their sample
    with |dDM/d$t$| > 0.01\,\DMunits\,yr$^{-1}$.
  }
  \label{DMcorrelations}
\end{figure*}

\cite{hlk+04} found a correlation between \ac{dm} and its time derivative in their sample of 374 pulsars. When fitting a power law, they found an exponent of 0.57(9), consistent with a square-root dependence: $|\text{dDM/d}t| = 0.0002\sqrt{\text{DM}}$ (DM in \DMunits{} and dDM/d$t$ in \DMunits\,yr$^{-1}$). Applying this analysis to our data, we get an exponent of 0.7(6), which is consistent with the \cite{hlk+04} result, albeit not very constraining. When we fit the amplitude of a square-root dependence, however, we get $|\text{dDM/d}t| = 0.000024(5)\sqrt{\text{DM}}$, highly inconsistent with the value from \cite{hlk+04}, which is an order of magnitude larger.
This suggests that this kind of analysis is biased. One major difference between the two studies is the sensitivity to \ac{dm}: Our single-observation measurements of DM are typically two orders of magnitude more precise than the overall \ac{dm} of \cite{hlk+04}, so we investigated possible correlations between DM, |dDM/d$t$|, and $\sigma_\text{DM}$ in both datasets by fitting power law relations (see Fig.~\ref{DMcorrelations}). Notably, the correlation between |dDM/d$t$| and $\sigma_\text{DM}$ in the \cite{hlk+04} dataset is nearly as strong as the correlation between |dDM/d$t$| and DM. Furthermore, only 21 out of the 374 sources show a |dDM/d$t$| measurement above $3\sigma$ significance, while two-thirds are below $1\sigma$, so a majority of their measurements are upper limits that can bias the fit.
Another potential bias lies in the source selection. The sample observed by \cite{hlk+04} is strongly focused on sources in the direction of the Galactic Centre. As the sensitivity of \ac{lofar} is strongly dependent on source elevation \citep[see Fig.~1 of][]{nsk+15}, the detectability of pulsars (and therefore our sample) is biased towards high-declination sources.

\begin{table}
  \caption{Quantities derived from the \ac{dm} time series of each pulsar.
    Given are the average \ac{dm} derivative over the entire dataset
    and the value of the structure function $D_\text{DM}$ at a time lag of 1000\,days.}
  \label{tableII}
  \centering
  \begin{tabular}{ c c c}
source~name   &  $\left<\frac{\text{dDM}}{\text{d}t}\right>$                     &  $D_\text{DM}(1000\,\text{d})$                          \\
(J2000)       &  $\left(10^{-4}\frac{\text{pc}}{\text{cm}^3\,\text{yr}}\right)$  &  $\left(10^{-6}\frac{\text{pc}^2}{\text{cm}^6}\right)$  \vspace{4pt}  \\  \hline  \rule{0pt}{15pt}\unskip
J0030+0451    &  \dcol{0.40(7)}                                                  &  \dcol{0.023}                                           \\
J0034$-$0534  &  \dcol{-0.37(2)}                                                 &  \dcol{0.016}                                           \\
J0218+4232    &  \dcol{-6.10(19)}                                                &  \dcol{6.2}                                             \\
J0407+1607    &  \dcol{-1.30(18)}                                                &  \dcol{0.56}                                            \\
J0621+1002    &  51(4)                                                           &  \dcol{220}                                             \\
J0645+5158    &  \dcol{0.02(4)}                                                  &  \dcol{0.021}                                           \\
J0740+6620    &  \dcol{4.40(5)}                                                  &  \dcol{0.83}                                            \\
J0751+1807    &  \dcol{-5.2(6)}                                                  &  \dcol{2.1}                                             \\
J0952$-$0607  &  \dcol{-0.6(2)}                                                  &  \dcol{0.17}                                            \\
J1012+5307    &  \dcol{1.30(3)}                                                  &  \dcol{0.086}                                           \\
J1022+1001    &  \dcol{0.02(5)}                                                  &  \dcol{0.0038}                                          \\
J1024$-$0719  &  \dcol{-1.7(4)}                                                  &  \dcol{0.90}                                            \\
J1125+7819    &  \dcol{1.70(5)}                                                  &  \dcol{0.30}                                            \\
J1300+1240    &  \dcol{-0.08(8)}                                                 &  \dcol{0.35}                                            \\
J1400$-$1431  &  \dcol{2.90(12)}                                                 &  \dcol{0.48}                                            \\
J1544+4937    &  \dcol{-0.6(5)}                                                  &  \dcol{0.046}                                           \\
J1552+5437    &  \dcol{-2.80(16)}                                                &  \dcol{0.50}                                            \\
J1640+2224    &  \dcol{0.42(4)}                                                  &  \dcol{0.038}                                           \\
J1658+3630    &  \dcol{1.7(2)}                                                   &  \dcol{0.44}                                            \\
J1713+0747    &  \dcol{-0.58(15)}                                                &  \dcol{0.052}                                           \\
J1730$-$2304  &  \dcol{3.5(5)}                                                   &  \dcol{1.9}                                             \\
J1738+0333    &  \dcol{-6.3(6)}                                                  &  \dcol{4.3}                                             \\
J1744$-$1134  &  \dcol{0.09(10)}                                                 &  \dcol{0.046}                                           \\
J1853+1303    &  \dcol{1.0(3)}                                                   &  \dcol{0.34}                                            \\
J1857+0943    &  \dcol{-1.7(6)}                                                  &  \dcol{1.5}                                             \\
J1911$-$1114  &  \dcol{-0.1(6)}                                                  &  \dcol{2.2}                                             \\
J1918$-$0642  &  \dcol{-1.6(14)}                                                 &  \dcol{2.7}                                             \\
J1923+2515    &  \dcol{2.00(11)}                                                 &  \dcol{0.33}                                            \\
J1944+0907    &  \dcol{-9.0(5)}                                                  &  \dcol{5.4}                                             \\
J1955+2908    &  -7(2)                                                           &  \dcol{23}                                              \\
J2043+1711    &  \dcol{0.90(9)}                                                  &  \dcol{0.23}                                            \\
J2051$-$0827  &  \dcol{-7.1(3)}                                                  &  \dcol{2.9}                                             \\
J2145$-$0750  &  \dcol{0.52(4)}                                                  &  \dcol{0.069}                                           \\
J2222$-$0137  &  \dcol{0.1(2)}                                                   &  \dcol{0.12}                                            \\
J2302+4442    &  \dcol{-9.0(4)}                                                  &  \dcol{6.2}                                             \\
J2317+1439    &  \dcol{-0.19(15)}                                                &  \dcol{1.3}                                             \\
\hline
\end{tabular}

\end{table}

\subsection{Structure functions}
\begin{figure*}
  \centering
  \includegraphics[width=\textwidth]{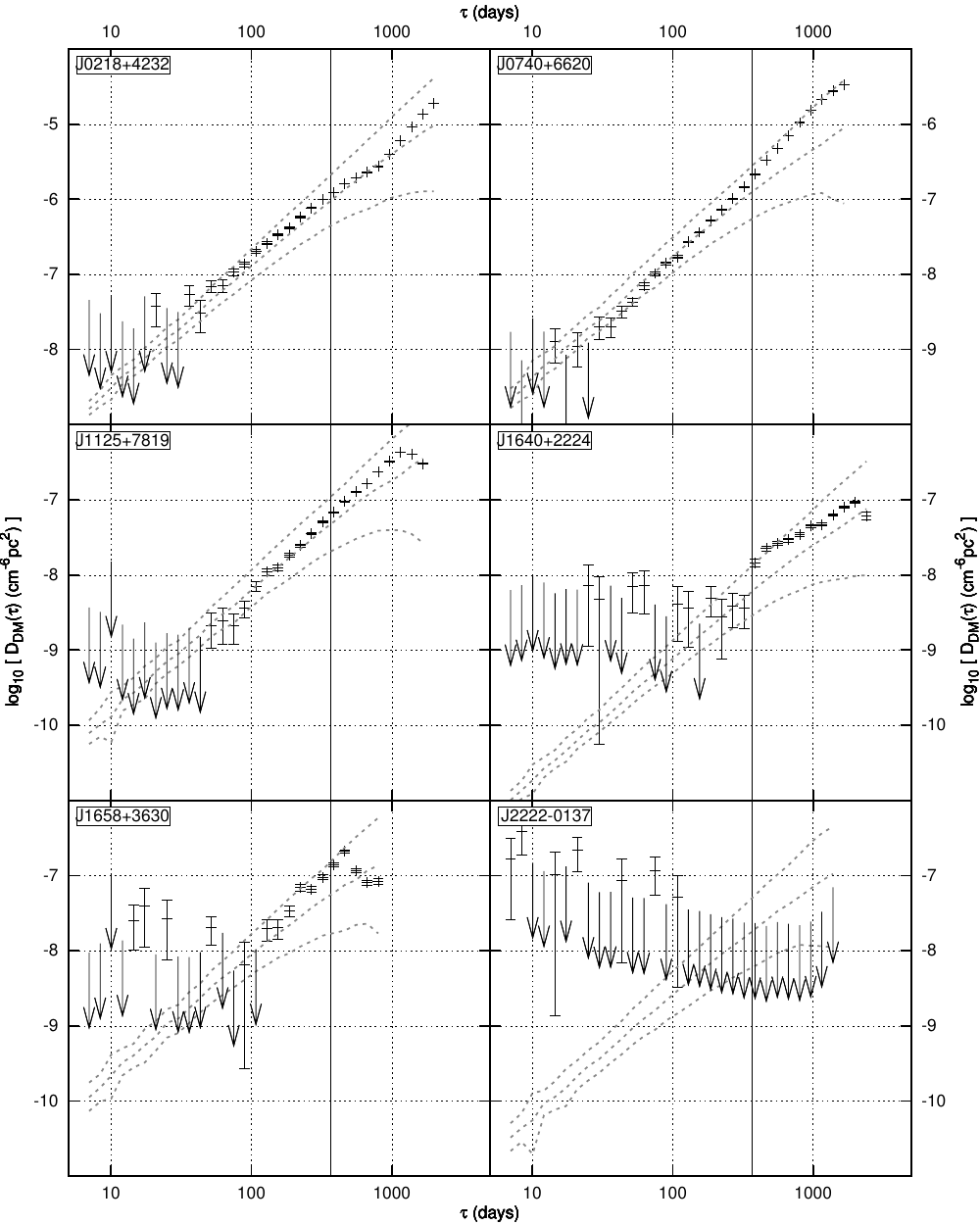}
  \caption{Structure functions for the \ac{dm} time series presented
    in Fig.~\ref{fig:DMvars_glow}.
    The dashed lines indicate the median and 68\% confidence interval
    of the Kolmogorov structure function simulations (see text for details).
    The solid black vertical line indicates $\tau = 1\,\text{yr}$.
    The white noise level has been subtracted
    from the data and the Kolmogorov model.
    Arrows indicate upper limits for
    measurements consistent with zero at the 1-$\sigma$ level.
  }
  \label{fig:sf_glow}
\end{figure*}

\begin{figure*}
  \centering
  \includegraphics[width=\textwidth]{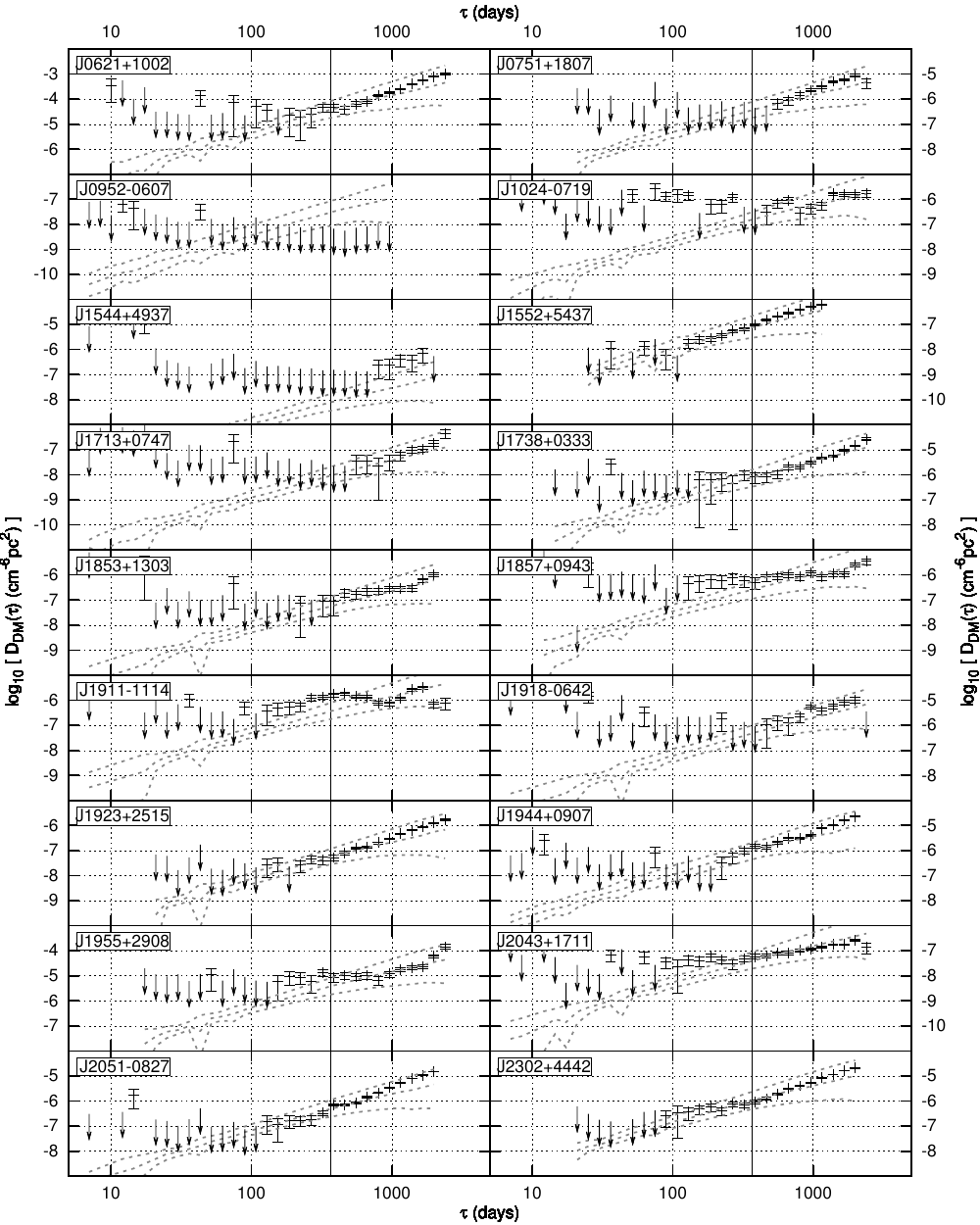}
  \caption{Same as Fig.~\ref{fig:sf_glow},
    but for the \ac{dm} time series presented
    in Fig.~\ref{fig:DMvars_core}.
  }
  \label{fig:sf_core}
\end{figure*}

The \ac{iism} turbulence is known to often follow a Kolmogorov spectrum \citep{ars95}.
To characterise the turbulence in our dataset, we investigated
the structure function of the \ac{dm} time series, $D_\text{DM}$, which gives rise to a statistical correlation between pairs of \acp{dm} at particular time lags $\tau$.
It is defined as follows \citep[see, e.g.][]{yhc+07}:
\begin{equation}
  D_\text{DM}(\tau) =
  \left< \left[ \text{DM}(t + \tau) - \text{DM}(t) \right]^2 \right>,
\end{equation}
where the angle brackets indicate the ensemble average.
For Kolmogorov turbulence, the structure function of the \ac{dm} is expected to be a power law of the form:
\begin{equation}
  D_\text{DM}(\tau) \propto \tau^{5/3}.
\end{equation}
In our analysis, we evaluated the structure function at equal intervals on a logarithmic scale. Figures~\ref{fig:sf_glow} and \ref{fig:sf_core} show the structure functions for the \ac{dm} time series presented in Figs.~\ref{fig:DMvars_glow} and \ref{fig:DMvars_core}, that is, all pulsars that are not significantly affected by the Solar wind. The amplitude of the Kolmogorov structure function fits (see below) at a time lag of 1000\,days is given in Table~\ref{tableII}.

To investigate whether our structure functions are compatible with the prediction arising from the assumption of a Kolmogorov spectrum, we simulated 1000 \ac{dm} time series per pulsar with the same sampling as our observations assuming a Kolmogorov power spectrum (i.e.\ with a spectral index of -8/3) and calculated the structure functions. From this set of simulated structure functions we took the median value at each time lag as well as the 68\% confidence interval.

To get an amplitude for the simulation, we first ran a least-squares routine to fit the amplitude of a Kolmogorov power law to the simulated structure function, weighting each sample by its inverse variance. Then we fit a Kolmogorov power law plus a constant (to account for the white noise) to the observed structure function, adding the relative uncertainties of the data and the simulations in quadrature to obtain a relative uncertainty to use in the fit. This choice of weights in the fit accounts for the high red-noise uncertainty at large lags, which is accounted for in our simulations but not in the data-derived structure function. Using the quotient of the two fit amplitudes, we re-scaled the simulated structure function to match the observed one.
We then subtracted the white-noise level from the observed structure function, such that the resulting structure function only contained the effects of turbulent processes.

\begin{figure*}
  \centering
  \includegraphics[width=\textwidth]{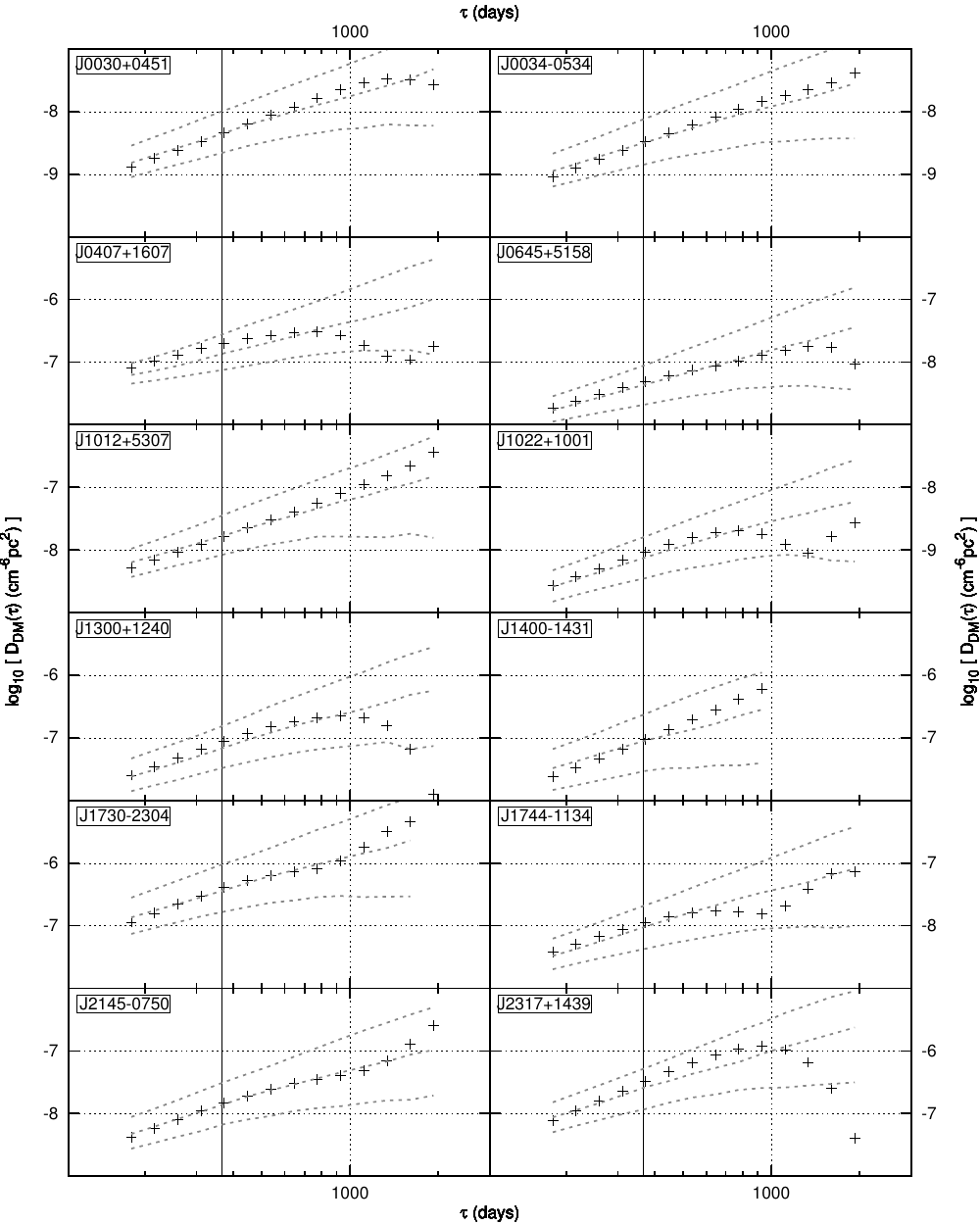}
  \caption{Structure functions for the \ac{dm} time series presented
    in Fig.~\ref{fig:DMvars_solar} after mitigation of the Solar wind signal.
    The representation is the same as in Fig.~\ref{fig:sf_glow}.
  }
  \label{fig:sf_solar}
\end{figure*}

In Figure~\ref{fig:sf_solar}, we present structure functions of the \ac{dm} time series that are affected by the Solar wind (Fig.~\ref{fig:DMvars_solar}). As we aim to quantify the \ac{iism} turbulence, we mitigated the effect of the Solar wind by sampling the \ac{dm} time series when the pulsar was furthest from the Sun. We then interpolated between those samples using a cubic spline. The structure function was only calculated for time lags greater than half a year as the short-term variations are underestimated from the interpolation.

From our simulations we conclude that all our structure functions are consistent with a Kolmogorov turbulence spectrum.
Many sources show a decrease (e.g.\ PSRs~J1300+1240 and J1125+7819) or dips (e.g.\ PSRs~J1022+1001 and J2145$-$0750) in the structure function at large lags. The reason behind this is that the large lags strongly depend on the start and end date of the observations, as those define a very specific realisation of the stochastic red-noise process. This is also reflected in the growing uncertainties of our simulations at large lags and can be seen in \citet[PSRs~J0437$-$4715 or J0711$-$6830]{kcs+13}.
For sources with clear linear trends in \ac{dm}, like PSRs~J0740+6620 or J1400$-$1431, the structure function tends to look steeper than Kolmogorov turbulence, while still being consistent with it. This is expected as the structure function of a linear gradient in \ac{dm} has a spectral index of 2 (compared to a spectral index of 5/3 for Kolmogorov turbulence). As discussed in Sect.~\ref{sec:dm_features}, such a linear gradient can be explained by the pulsars' motion along the line of sight \citep{lcc+16}.
PSR~J2317+1439 shows an abrupt reversal in the slope of the \ac{dm} time series (see Fig.~\ref{fig:DMvars_solar}), which is unusual for a Kolmogorov process. Further monitoring of this source will reveal whether the structure function will remain consistent with Kolmogorov turbulence for time lags above $\sim$1000\,days.

\subsection{Comparison to PTA results}
As the mitigation of the \ac{iism} is an important consideration in \ac{pta} gravitational-wave analysis, \ac{dm} time series of \ac{pta} pulsars are studied extensively \citep[e.g.][]{kcs+13,dcl+16,jml+17}, although these use much lower cadence and are obtained at much higher frequencies (typically $\sim$300-2000\,MHz).
The published \ac{pta} \ac{dm} time series present datasets that mostly end around the start of our dataset, making a direct comparison of the \ac{dm} time series difficult.

\cite{kcs+13} used observations from three frequency bands centred at $\sim$700\,MHz, $\sim$1400\,MHz and $\sim$3100\,MHz. From these data, they derived one \ac{dm} estimate every three months for the 20 pulsars in their sample. As this work used the \ac{ppta} observations from the Parkes telescope (which is located in the southern hemisphere), there are only seven pulsars from their study that we also observed -- all of which are at a low declination where the sensitivity of \ac{lofar} decreases dramatically \citep[e.g.][]{nsk+15}.
\cite{kcs+13} do not see clear deviations from the simple symmetric Solar wind model they applied, but this is expected as their \ac{dm} precision is not high enough. However, for some pulsars (e.g.\ PSR~J1857+0943), their \ac{dm} precision is comparable to the one achieved with \ac{lofar}.
\cite{kcs+13} also give the value of the structure function at a time lag of 1000\,days, which is consistent with our results within the red-noise uncertainty.

\cite{dcl+16} present timing results for 42 \acp{msp} observed with telescopes of the \ac{epta}. The observations were taken across a range of frequencies between 350\,MHz and 2.6\,GHz, with the majority being taken around 1.4\,GHz. Of the pulsars in their study, 20 are in common with this work.
In their paper, they do not present \ac{dm} time series, but instead used the \textsc{temponest} software to model the \ac{dm} variations as a second-order polynomial plus a spectral noise model. This makes a comparison of individual measurements difficult, but the large-scale trends can be compared.
Whether or not the \ac{lofar} data provide more information on the \ac{dm} variability is again strongly pulsar dependent. For example, we detect a very clear \ac{dm} signal in PSR~J2317+1439 (see Fig.~\ref{fig:DMvars_solar}) while \cite{dcl+16} only find a \ac{dm} trend at the $1\sigma$-level, consistent with the overall trend we see in the early part of our dataset.

\cite{jml+17} present data from the \ac{nanograv} collaboration, analysing the \ac{dm} time evolution of 37 \acp{msp} at frequencies between 300\,MHz and 2.4\,GHz, 18 of which we analysed in this study. Again, the question of which instrument provides greater precision in \ac{dm} measurements is pulsar dependent: \ac{lofar} provides a much higher precision for PSR~J1012+5307, whereas the higher-frequency \ac{nanograv} data provide a much higher precision for PSR~J1857+0943 \citep[PSR~B1855+09 in][]{jml+17}.

Overall, we find no inconsistency between our results and the highlighted recent \ac{pta} publications. However, this is partly caused by the fact that the datasets are difficult to compare due to a lack of overlap in observing epochs, and differences in data representation.

\subsection{Consequences for PTAs}
In pulsar timing experiments, the actual \ac{toa} is usually not the measure of interest. Instead, the \ac{dm}-corrected, infinite-frequency \ac{toa} $T_\infty$ is the relevant measure when trying to measure \ac{iism}-independent effects. As variations in the \ac{dm} at a relevant magnitude are very common, they usually have to be accounted for \citep{vlh+16}. In the following, we will discuss different approaches to apply corrections of the dispersive delays to \acp{toa} at 1.4\,GHz, with a timing precision of $\sigma_\text{ToA} = \text{\SI{1}{\micro\second}}$.

There are two basic approaches to consider. Approach~I would be to calculate the in-band \ac{dm} (as in this paper) and apply the according dispersive delay (using Eq.~\ref{eq:disp_delay}) to the timing \acp{toa}. The uncertainty on $T_\infty$ would then be
\begin{equation}
  \sigma_\infty = \sqrt{\sigma_\text{ToA}^2 + \sigma_\text{disp}^2}.
\end{equation}
To avoid the dispersive correction of being the dominant factor in this example ($\sigma_\text{ToA} = \text{\SI{1}{\micro\second}}$ at 1.4\,GHz), the uncertainty on the \ac{dm} would have to be lower than $5\times10^{-4}$\,\DMunits, which would not be achievable from in-band measurements at 1.4\,GHz with a bandwidth of $\sim$300\,MHz, but potentially from simultaneous lower-frequency \ac{dm} measurements as presented in this paper. Additionally, high-cadence \ac{dm} time series could be smoothed to increase the precision of the dispersive corrections if the \ac{dm} variations are smooth.

Approach~II would be to use \acp{toa} from multiple frequency bands and fit for the \ac{dm} and $T_\infty$ simultaneously. This has the advantage of using the entire available bandwidth, but the disadvantage that exactly simultaneous observations may not be available, in which case short-term signatures from any timing parameter (e.g.\ parameters describing the binary motion) have to be ruled out. Also, non-simultaneous observations can have a significantly different \ac{dm} if they are too far separated in time \citep[see][]{lccd15}.

\begin{figure}
  \hspace{-10pt}\includegraphics[width=.48\textwidth]{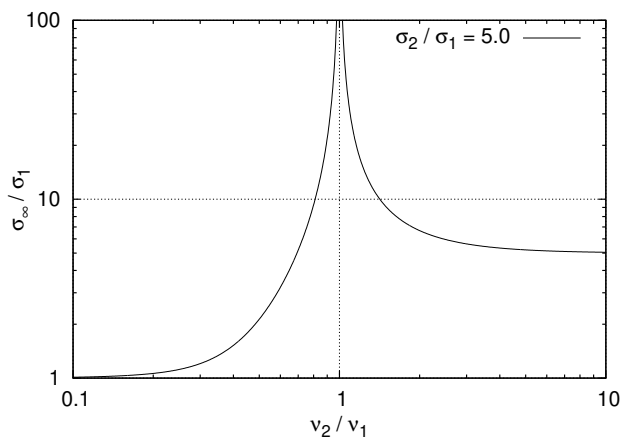}
  \caption{Illustration of Eq.~\ref{eq:sig_tinf} \citep[Eq.~12 in][]{lbj+14}
    for an observation at two frequency bands.
    The uncertainty of the infinite-frequency \ac{toa} is plotted
    as a function of $\nu_2$, with $\nu_2$ providing
    less precise \acp{toa} than $\nu_1$ ($\sigma_2=5.0\sigma_1$).
    All quantities are given relative to the frequency
    and \ac{toa} uncertainty of Band~1.
    On the left side of the plot ($\nu_2 \ll \nu_1$), 
    $\sigma_\infty$ approaches $\sigma_1$,
    whereas on the right side ($\nu_2 \gg \nu_1$)
    it approaches $\sigma_2$.
    This shows that the precision of the higher-frequency band
    is the limiting factor on the precision of the infinite-frequency \ac{toa}.
  }
  \label{fig:sig_tinf_vs_nu2}
\end{figure}

\cite{lbj+14} investigated how the choice of the observing frequencies $\nu_i$ and the corresponding \ac{toa} uncertainties $\sigma_i$ affect the precision of the infinite-frequency \acp{toa}. Equation~12 of their paper shows this relation in the case of only two observing frequencies:
\begin{equation}
  \left< \delta T_\infty^2 \right> =
  \frac{\sigma_1^2\nu_1^4 + \sigma_2^2\nu_2^4}{(\nu_1^2 - \nu_2^2)^2} .
  \label{eq:sig_tinf}
\end{equation}
The $\sigma_\infty$ can be computed as the square root of the above expression.
While the special case of only two \acp{toa} is simplified, the general conclusions also hold for cases with additional observing frequencies.
Most importantly, the frequency band with the largest $Q = \sigma^2\nu^4$ dominates the uncertainty in $T_\infty$. Due to the strong frequency dependence of $Q$, the highest frequency has to be the most-precisely measured in order to benefit from the additional bandwidth.
If $\nu_1$ is our most precise timing frequency and observations at $\nu_1$ and $\nu_2$ are used to calculate $T_\infty$ ($\sigma_2 > \sigma_1$), Fig.~\ref{fig:sig_tinf_vs_nu2} illustrates that $\nu_2$ should be smaller than $\nu_1$ to achieve a timing precision close to $\sigma_1$. If $\nu_2$ is the higher frequency, $\sigma_\infty$ is limited by $\sigma_2$.
In \ac{pta} data, the smallest \ac{toa} uncertainty is often obtained at 1.4\,GHz, so observations at lower frequencies should be used to correct for \ac{dm} variability. In the case of the \ac{lofar} observations presented in this paper, a frequency-integrated \ac{toa} uncertainty of $\sim$\SI{100}{\micro\second} or better would be required to not dominate \ac{dm} corrections at the \SI{1}{\micro\second}-level. This condition is fulfilled for all pulsars in our sample except for PSR~J0621+1001 (see Table~\ref{obs_table}), which is a pulsar that is usually timed at an RMS much worse than \SI{1}{\micro\second} \citep[see e.g.][]{dcl+16}. This shows that Approach~II should be preferred over Approach~I, if it is applicable.

To illustrate the superiority of Approach~II over Approach~I, we will go through an example of correcting for the dispersive delay in 1.4\,GHz observations using \ac{lofar}. We assume $\nu_1 = 1.4\,\text{GHz}$, $\sigma_1 = \text{\SI{1}{\micro\second}}$, $\nu_2 = 150\,\text{MHz}$ and $\sigma_2 = \text{\SI{100}{\micro\second}}$. We note that a \ac{toa} precision of \SI{100}{\micro\second} is a worst-case scenario. Using Approach~II, we get $\sigma_\infty = \text{\SI{1.5}{\micro\second}}$. For Approach~I, we assume ten \acp{toa} around $\nu_2$ across a bandwidth of 75\,MHz, each with an uncertainty of $\sqrt{10}\times\text{\SI{100}{\micro\second}}$ and use Eq.~10 of \cite{lbj+14} to compute the \ac{dm} uncertainty. The resulting $\sigma_\infty$ is \SI{3.8}{\micro\second}, which is worse by a factor of 2.5.

A major caveat of using \ac{lofar} observations for \ac{dm} corrections at higher frequencies lies in the spectra of pulsars: Pulsars with steep spectra are bright at low frequencies and allow for highly precise \ac{dm} corrections, but are fainter at high frequencies, such that the timing precision may be so low that \ac{dm} corrections are less relevant. On the other hand, pulsars with flatter spectra tend to have a high \ac{toa} precision at high frequencies, but are very faint at low frequencies due to the steep spectral index of the sky background noise, which is of order 2.5 \citep{lmop87}.
For example, the pulsar with the best \ac{dm} precision in our sample is PSR~J0034$-$0534, but its timing RMS at 1.4\,GHz is only average \citep[\SI{4.27}{\micro\second}, see][]{pdd+19}.
The well-timed PSR~J1713+0747, however, is so faint we cannot even detect it in \ac{glow} data and achieve a \ac{dm} precision an order of magnitude worse than for PSR~J0034$-$0534.

It therefore follows that Approach~I using \ac{lofar} data would currently not improve the \ac{pta} timing precision for most sources, which could be solved by increasing the timing precision of steep-spectrum sources at 1.4\,GHz or improving the \ac{dm} precision of flat-spectrum sources at low frequencies. The latter could be achieved by greatly increasing the integration time with the \ac{lofar} Core or using more sensitive telescopes like the upcoming Square Kilometre Array phase-one low-frequency bands \citep[SKA1-LOW; e.g.\ ][]{jhm+15}. Using a slightly higher frequency might also help if the spectrum of the pulsar is very flat -- the optimal frequency for the \ac{dm} measurement is strongly pulsar dependent.
As the timing precision of steep-spectrum \ac{pta} sources continues to increase due to improved telescopes, the significance of highly precise \ac{dm} time series such as those presented in this work will become crucial for \ac{dm} corrections.
However, Approach~II should already now improve the timing precision of \acp{pta} for many sources, as it provides more precise corrections for the dispersive delays. We reserve such an analysis for a future paper, however.

\subsection{Data access}
Our \acp{toa}, timing models, templates and \ac{dm} time series are available online on Zenodo\footnote{\url{https://zenodo.org/deposit/4290012} \\ DOI: 10.5281/zenodo.4290012}.
The raw observations taken with the \ac{lofar} Core can be accessed via the \ac{lofar} Long Term Archive\footnote{\url{https://lta.lofar.eu/}}. The raw \ac{glow} data are available upon request.

\section{Conclusions}
\label{sec:concl}
We present low-frequency \ac{dm} time series for \npsr{} \acp{msp} over up to \maxspan{} years, obtained from observations with the \ac{lofar} Core and the individual \ac{glow} telescopes. Except for the pulsars with very high \ac{dm} uncertainty (i.e.\ greater than $\sim10^{-4}$\,\DMunits{}), all pulsars show significant variations in \ac{dm}. Twelve pulsars show a clear Solar wind signal in their \ac{dm} time series that can usually not be modelled by a spherically symmetric electron content with constant amplitude, which is often used in pulsar timing. All of the \ac{iism}-related \ac{dm} variations we present are consistent with a Kolmogorov turbulence spectrum.


Acknowledging the caveat that \ac{lofar} often provides a high \ac{dm} precision for pulsars that are poorly timed at higher frequencies, and vice versa, we show that our \ac{lofar} \ac{dm} monitoring could be used to correct variations of the dispersive delays in higher-frequency observations from \acp{pta}. We do not find evidence for a frequency-dependent \ac{dm}, so we expect the impact of this effect to be limited.

\begin{acknowledgements}
We like to thank W.~A.~Coles and M.~Lam for useful discussions.
JPWV acknowledges support by the Deutsche Forschungsgemeinschaft (DFG) through the Heisenberg programme (Project No. 433075039).
CT is a Veni fellow (project number 016.Veni.192.086), partly financed by the Dutch Re-search Council (NWO).
RPB acknowledges support of the European Research Council, under the European Union's Horizon 2020 research and innovation program (grant agreement No. 715051; Spiders).
JvL acknowledges funding from the European Research Council under the European Union's Seventh Framework Programme (FP/2007-2013) / ERC Grant Agreement n. 617199 (`ALERT'), and from Vici research programme `ARGO' with project number 639.043.815, financed by the Dutch Research Council (NWO).

Part of this work is based on data obtained with the International LOFAR Telescope under project codes LC0\_011, LC1\_027, LC2\_010, LT3\_001, LC4\_004, LT5\_003, LC9\_041 and LT10\_004. LOFAR \citep{vwg+13} is the Low Frequency Array designed and constructed by ASTRON. It has observing, data processing, and data storage facilities in several countries, that are owned by various parties (each with their own funding sources), and that are collectively operated by the ILT foundation under a joint scientific policy. The ILT resources have benefitted from the following recent major funding sources: CNRS-INSU, Observatoire de Paris and Universit\'{e} d'Orl\'{e}ans, France; BMBF, MIWF-NRW, MPG, Germany; Science Foundation Ireland (SFI), Department of Business, Enterprise and Innovation (DBEI), Ireland; NWO, The Netherlands; The Science and Technology Facilities Council, UK.

This paper uses data obtained with the German LOFAR stations,
during station-owners time and ILT time allocated under project codes
LC0\_014, LC1\_048, LC2\_011, LC3\_029, LC4\_025, LT5\_001, LC9\_039 and LT10\_014.
We made use of data from
the Effelsberg (DE601) LOFAR station funded by the Max-Planck-Gesellschaft;
the Unterweilenbach (DE602) LOFAR station funded by
the Max-Planck-Institut für Astrophysik, Garching;
the Tautenburg (DE603) LOFAR station funded by the State of Thuringia,
supported by the European Union (EFRE) and the
Federal Ministry of Education and Research (BMBF) Verbundforschung
project D-LOFAR I (grant 05A08ST1);
the Potsdam (DE604) LOFAR station funded by the
Leibniz-Institut für Astrophysik, Potsdam;
the Jülich (DE605) LOFAR station supported by the
BMBF Verbundforschung project DLOFAR I (grant 05A08LJ1);
and the Norderstedt (DE609) LOFAR station funded by the
BMBF Verbundforschung project D-LOFAR II (grant 05A11LJ1).
The observations of the German LOFAR stations
were carried out in stand-alone GLOW mode,
which is technically operated and supported by
the Max-Planck-Institut für Radioastronomie, the Forschungszentrum
Jülich and Bielefeld University. We acknowledge support and
operation of the GLOW network, computing and storage facilities by
the FZ-Jülich, the MPIfR and Bielefeld University and financial support
from BMBF D-LOFAR III (grant 05A14PBA) and D-LOFAR IV (grant 05A17PBA),
and by the states of Nordrhein-Westfalia and Hamburg.
We acknowledge the work of A.~Horneffer in setting up
the GLOW network and initial recording machines.
\end{acknowledgements}

\bibliographystyle{aa}
\bibliography{journals,psrrefs,modrefs,crossrefs,additions}

\end{document}